\documentclass[a4paper]{article}

\usepackage{INTERSPEECH2021}
\usepackage{multirow}
\usepackage{tipa}
\usepackage{mwe}
\usepackage{subfigure}
\title{Detection of Consonant Errors in Disordered Speech Based on Consonant-Vowel Segment Embedding}
\name{Si-Ioi Ng$^1$, Cymie Wing-Yee Ng$^2$, Jingyu Li$^1$, Tan Lee$^1$}
\address{
  $^1$Department of Electronic Engineering, The Chinese University of Hong Kong \\
  $^2$Department of Otorhinolaryngology, Head \&  Neck Surgery, The Chinese University of Hong Kong}
\email{\{siioing, cymieng, lijingyu0125\}@link.cuhk.edu.hk, tanlee@ee.cuhk.edu.hk}

\begin{document}

\maketitle
\begin{abstract}

Speech sound disorder (SSD) refers to a type of developmental disorder in young children  who encounter persistent difficulties in producing certain speech sounds at the expected age. Consonant errors are the major indicator of SSD in clinical assessment. Previous studies on automatic assessment of SSD revealed that detection of speech errors concerning short and transitory consonants is less satisfactory. This paper investigates a neural network based approach to detecting consonant errors in disordered speech using consonant-vowel (CV) diphone segment in comparison to using consonant monophone segment. The underlying assumption is that the vowel part of a CV segment carries important information of co-articulation from the consonant. Speech embeddings are extracted from CV segments by a recurrent neural network model. The similarity scores between the embeddings of the test segment and the reference segments are computed to determine if the test segment is the expected consonant or not.  Experimental results show that using CV segments achieves improved performance on detecting speech errors concerning those “difficult” consonants reported in the previous studies.




\end{abstract}

\section{Introduction}\label{intro}


In the process of language acquisition, children are expected to master the language's speech sounds in stages and be able to self-correct mistakes when growing up. Yet, a significant percentage of children may encounter persistent difficulties in producing certain sounds correctly after the expected age of acquisition.
These children are likely to be diagnosed as having speech sound disorder (SSD).
If left untreated, the children would face significant long-term challenges in education and social life \cite{hitchcock2015social}. 
Early-stage diagnosis is therefore important for effective intervention and rehabilitation \cite{rvachew2014report}. Traditionally clinical diagnosis of SSD is carried out by qualified speech and language pathologists (SLPs).
With the long-lasting shortage of and increasing demand for SLPs, automated detection of speech disorder
is a highly desirable approach to providing timely assessment and/or screening of large population of children.

Detection of SSD is formulated as a task of distinguishing disordered speech sounds from typical ones based on acoustic speech signals. 
Many studies have been focused on consonant error detection, given that assessment of consonant pronunciation is a major task in clinical diagnosis of SSD. 
In \cite{shahin2014comparison,ward2016automated}, constrained lattice was incorporated from an automatic speech recognition (ASR) system, by which expected ASR outputs could be preset to facilitate the detection of target consonants. In \cite{dudy2018automatic}, goodness of pronunciation (GOP) was used to detect the deviation of pronunciations in disordered speech. The GOP was computed based on the likelihood ratio of the expected consonant versus other phonemes.
In \cite{yeung2017predicting}, child speech was evaluated by template matching, where the test segment was compared with reference segments by cosine distance.

In our recent work \cite{Ng2020}, Siamese auto-encoder network was applied to contrast hypothetically disordered consonant segments against typical ones. It was found that 
the detection performance on un-aspirated stop consonants was consistently less satisfactory than on other consonants. Similar results were reported in Wang et al. \cite{Wang2019}. As an un-aspirated stop consonant is preceded by a long closure (caused by blockage of air flow), the detection is likely to be interfered by background noise and hence becomes unreliable.
\begin{figure}[t!]
\centering
\subfigure[]{
\includegraphics[width=3.8cm]{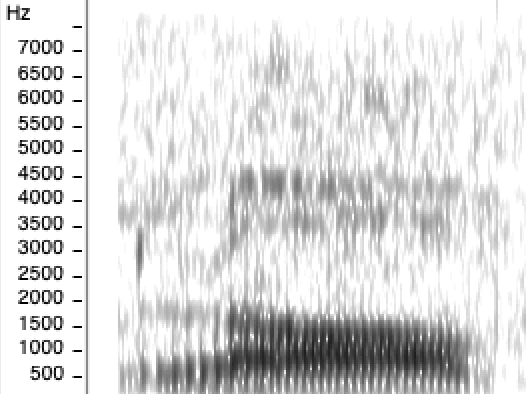}
}
\subfigure[]{
\includegraphics[width=3.8cm]{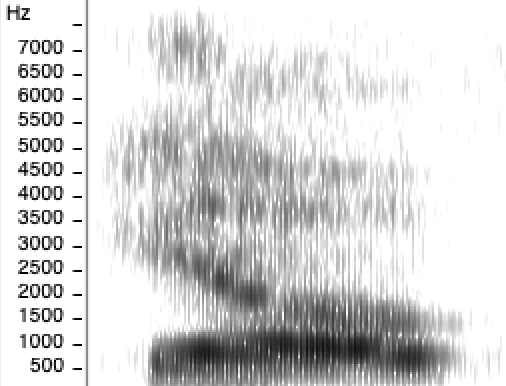}
}
\caption{
Comparison of Cantonese syllable /\textipa{l O:}/ uttered by
(a) a typically developing speaker; (b) a disordered speaker making a gliding error by substituting /l/ with /j/.}
\label{fig:gliding_demo}
\end{figure}
From a different perspective, contrastive characteristics of consonants are also carried by the neighbouring vowels, as speech sounds are co-articulated instead of produced discretely \cite{lee2017cv}.
Given that substitution of consonants is a prominent indicator of SSD, the co-articulated vowel segments are expected to exhibit atypical acoustic features when SSD occurs. Figure \ref{fig:gliding_demo} illustrates the impact of SSD, where the expected consonant /l/ is substituted by /j/ in the disordered speech. The substituted consonant alters formant transition in the vowel segment /\textipa{O:}/. Similar observations were reported in \cite{lindblom+1963, Stevens+1963}. It was suggested that the acoustic realization of vowels was affected by the place, manner and voicing characteristics of neighboring consonants \cite{lindblom+1963, Stevens+1963}. 


The significance of contextual effect motivates us to 
investigate the use of consonant-vowel (CV) di-phone segments for SSD detection in child speech. A common type of SSD can be presented as the desired consonant being substituted by another consonant. In the present study, detection of consonant errors in child speech is formulated as a binary classification problem to distinguish the desired consonant from other consonants based on mono-phone segments and/or di-phone segments.

\section{Background Knowledge about Cantonese}
\begin{figure}[h!]
  \setlength\belowcaptionskip{-0.8\baselineskip}
  \centering
  \includegraphics[width=\linewidth]{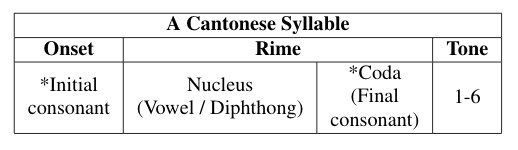}
  \caption{Structure of Cantonese syllable. '*' denotes optional.} 
  \label{fig:syllable_structure}
\end{figure}

This study is focused on Cantonese. It is a major Chinese dialect widely spoken in Hong Kong, Macau, Guangdong and Guangxi Provinces of Mainland China, as well as overseas Chinese communities. 
Cantonese is a monosyllabic and tonal language. Each Chinese character is pronounced as a single syllable carrying a lexical tone. As illustrated in Figure \ref{fig:syllable_structure},
a legitimate Cantonese syllable can be divided into an onset and a rime. The onset is a consonant. The rime can contain a nucleus or a nucleus followed by a coda, where the nucleus can be a vowel or a diphthong, while the following coda can be a final consonant. Both the onset and the coda are optional in a Cantonese syllable. For syllables without an initial consonant, it can be regarded as a null initial. 
In Cantonese, there are a total of $19$ initial consonants, $11$ vowels, $11$ diphthongs, $6$ final consonants and $6$ distinct lexical tones \cite{lee2002spoken,bauer2011modern}. In the present study we focus on speech segments that contain an initial consonant followed by a vowel nucleus.

\section{SSD Detection System}

\begin{figure}[t]
  \setlength\belowcaptionskip{-0.8\baselineskip}
  \centering
  \includegraphics[width=0.95\linewidth]{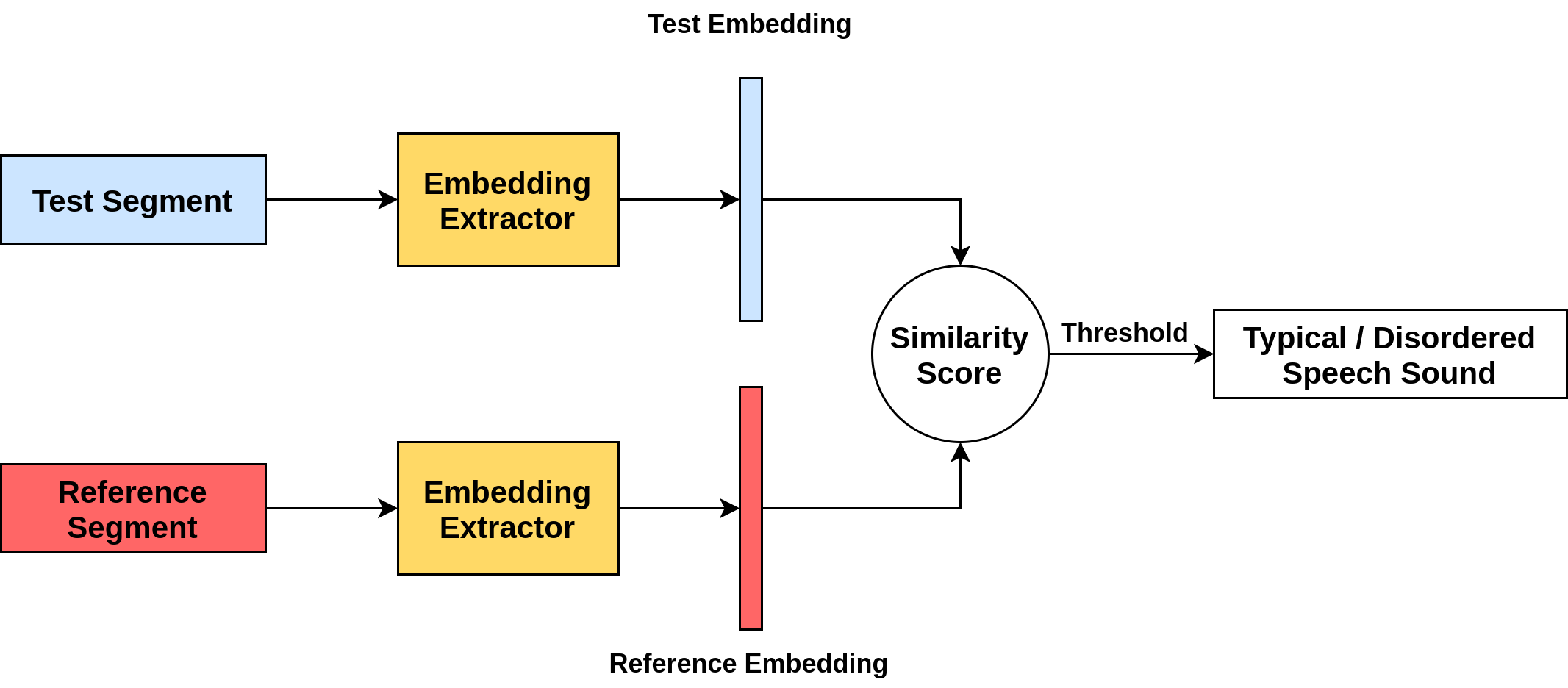}
  \caption{Speech sound disorder (SSD) detection system.} 
  \label{fig:system_design}
\end{figure}

The design of detection system follows how SLPs perceive consonant errors in disordered speech.
In the articulation test for SSD assessment, the SLPs decide whether the speech production of the desired consonant deviates from the typical ones. 
When the mismatch between the expected and the actual speech sound occurs, a phoneme, typically a consonant which is best described to substitute the expected consonant is used to annotate the consonant error. 
The classification of the error patterns is performed by the SLPs in the analysis of the assessment results.
For instance, \textipa{/p\textsuperscript{h}/} substituted by \textipa{/p/} is classified as a de-aspiration error, while \textipa{/s/} articulated as \textipa{/ts/} is an affrication error, while a distorted /s/ is a distortion error, etc. In the present study, we focus on a set of initial consonants in Cantonese, which are deemed indicative of child speech acquisition.


Towards the automatic assessment of SSD, the proposed system aims to determine if a consonant error occurs in the test speech segment.
As illustrated in Figure \ref{fig:system_design}, the test segment contains a specific consonant (C), or a consonant-vowel (CV) concatenation as part of the test word.
For consonant error detection, the test segment is compared, in a pairwise manner, with one or multiple reference segments that represent the expected phone. The reference segments are obtained from the speech of typically developing (TD) speakers.
The comparison between a pair of segments is carried out based on speech embedding, which is a fixed-dimension representation that encodes  acoustic information of speech segments into a low-dimension space. It allows flexible modeling and processing of speech segments of variable length for different downstream tasks, e.g., spoken term detection and discovery \cite{kamper2021improved}, pathological speech classification \cite{botelho2020pathological}, prediction of speech intelligibility score \cite{quintas2020automatic}, etc.
A similarity or distance score can be calculated on each pair of embeddings. Using a pre-defined threshold, a test segment can be classified as typical or disorder speech.


The embedding extractor is a trainable bidirectional gated recurrent unit (Bi-GRU) \cite{GRU2014}. The Bi-GRU is able to handle variable-length speech input with a simpler network structure.
Two extractors are trained separately for mono-phone C segments or diphone CV segments. The training is done with a multi-task approach.
The tasks comprises a softmax-based multi-class classification task and a binary classification task with cross-entropy objective functions.
The design of the extractors was motivated by the relation classifier used in computer vision and speaker verification systems \cite{xie2018comparator, nagrani2020voxceleb, Li2020}, and the inter- and intra-class relationship between speech segments are considered concurrently.
Let $\boldsymbol{x_t}$ and $\boldsymbol{x_r}$ denote the embedding of test segment and that of reference segment respectively. The cosine similarity score and the binary relation score are given as,
\begin{equation} \label{score_cos}
    score_{cos} = \frac{\boldsymbol{x_t}\cdot\boldsymbol{x_r}}{\lVert\boldsymbol{x_t}\rVert\lVert\boldsymbol{x_r}\rVert}
\end{equation}

\begin{equation} \label{score_binary}
\begin{split}
    score_{binary} = \sigma(\boldsymbol{W}((\boldsymbol{x_t} - \boldsymbol{x_r})\otimes(\boldsymbol{x_t} - \boldsymbol{x_r}))+\boldsymbol{b})
\end{split}
\end{equation}
where $\boldsymbol{W}$ and $\boldsymbol{b}$ are the weight matrix and bias that transform the embedding to binary relation score. $\sigma$ denotes the Sigmoid function. The two scores are combined as,
\begin{equation} \label{score_combined}
\begin{split}
    score_{C/CV} = \lambda * score_{cos} + (1-\lambda) * score_{binary}
\end{split}
\end{equation}
where the scalar weight $\lambda \in [0, 1]$. 
For C and CV segments respectively, the combined scores are denoted as $score_{C}$ and $score_{CV}$. The two scores can be fused to leverage the complimentary information from mono-phone and di-phone segments, i.e.,

\begin{equation} \label{score_final}
\begin{split}
    score_{FINAL} = w * score_{C} + (1-w) * score_{CV}
\end{split}
\end{equation}
where the weight $w \in [0, 1]$. Both $\lambda$ and $w$ are determined experimentally.


\section{Experimental Setup}

\subsection{Speech data}


Experiments on consonant error detection in child speech are carried out with a large-scale speech database named CUCHILD \cite{Ng2020CUCHILD}. CUCHILD was developed to support the research on automatic assessment of SSD as well as clinical studies of SSD in Cantonese-speaking children. It contains speech data from $1,986$ kindergarten children aged $3$ to $6$ whose first language is Cantonese.
All child subjects were formally assessed with the Hong Kong Cantonese Articulation Test (HKCAT) \cite{cheung2006hong}. As a result, $230$ children in the database were found to have SSD. Speech recordings were made in a number of kindergartens in Hong Kong. A digital recorder was placed at $20$-$50$ centimeters in front of the child. Each child was guided to produce a total of $130$ words in Cantonese. The word length is 1 to 4 syllables.
Detailed data processing and annotation work on the entire CUCHILD database are ongoing. In the present study, a subset of speech data from 172 TD children and 31 children with SSD are used in the experiment. All audio data are sampled at 16kHz and represented by 16-bit PCM.

In addition to CUCHILD, a large-vocabulary database of adult speech, named CUSENT, was used to provide additional training data for the proposed detection system. CUSENT contains about 20 hours of speech, with $20,000$ utterances from $76$ adult speakers \cite{lee2002spoken}.



\begin{table}[t!]
\caption{Number of C / CV segments for training and test. }
\centering
\resizebox{0.97\linewidth}{!}{%

\begin{tabular}{|c|c|c|}
\hline
\textbf{Name of subset}         & \textbf{\begin{tabular}[c]{@{}c@{}} No. of \\ segments (C/CV) \end{tabular}} & \textbf{\begin{tabular}[c]{@{}c@{}} No. of \\ speakers \end{tabular}} \\ \hline
\begin{tabular}[c]{@{}c@{}}Training \\ (Child, speed augmented)\end{tabular}       & 84,842  &                  153                                           \\ \hline
Training (Adult)       & 207,190    &   68                                                       \\ \hline
Test (Child, TD)       & 3,664   &      19                                                       \\ \hline
Test (Child. Atypical) & 1,384     &    31                                                       \\ \hline
\end{tabular}
}
\label{tab:numsegments}
\end{table}

\subsection{Data pre-processing}

C and CV speech segments are extracted from TD, atypical child speech and adult speech by forced alignment using GMM-HMM triphone acoustic models (AM). The AMs of child speech and adult speech are separately trained by the Kaldi speech recognition toolkit \cite{povey2011kaldi}.

The child speech model is trained on speech data from $153$ TD children of age $4$ to $6$. 
The acoustic features for GMM-HMM training consist of $13$-dimensional Mel-frequency cepstral coefficients (MFCC) and their first- and second-order derivatives extracted at every $0.003$ second. The choice of step size aims to obtain more precise alignment of C and CV speech segments. 
Linear discriminant analysis (LDA), semi-tied covariance (STC) transform and feature space Maximum Likelihood Linear Regression (fMLLR) are applied to the triphone AM training \cite{duda2012pattern, gales1999semi, gales1998maximum}. 
The training of adult AM follows the same recipe using speech data from $68$ adults in CUSENT. 
The child AM achieves a syllable error rate (SER) of $25.53$\% in free-loop syllable recognition of test speech from $19$ TD children.  
An SER of $11.66\%$ is obtained from the adult AM in the recognition of test speech from $8$ adults using a bi-gram language model. 

Speed augmentation is applied to increases the amount of child speech training data by 3-fold. This is done by altering the speed to 90\% and 110\% of the original speech rate \cite{ko2015audio}. 
The number of speech segments obtained by forced alignment, and the number of speakers used in the experiment, are listed as in Table \ref{tab:numsegments}. 

\subsection{Training of embedding extractors}
The Bi-GRU in each extractor consists of 3 hidden layers, $400$ hidden units in each layer. An embedding of $128$ dimensions is extracted from the last hidden layer of Bi-GRU with a fully-connected layer. The networks are built using PyTorch \cite{paszke2019pytorch} and are trained by the Adam optimizer \cite{kingma2014adam}. The joint objective function is composed by a multi-class cross-entropy and a binary cross-entropy. The cross-entropy is computed on $19$ output targets in the C embedding extractor and $173$ in the CV embedding extractor. These are defined in accordance to the number of initial consonants in Cantonese and the consonant-vowel combinations in the corpora.
The binary cross-entropy is calculated by randomly pairing with 4 speech segment for each training sample.
The network training setup includes a batch size of $256$, a learning rate of $0.001$, a dropout rate of $0.5$, and a weight decay of $0.0005$.  
The inputs to the Bi-GRUs are 80 dimensional Filter-bank features with global mean and variance normalization. The epoch of network training is empirically set to 5
to prevent overfitting.

\subsection{Evaluation metrics}
The detection performance are evaluated by two metrics, namely the equal error rate (EER) and the area under curve (AUC).
The EER represents the point where the false positive rate (FPR) equals to the false negative rate (FNR). It has been widely used in the biometetric security system.  
We define a pair of input segments to be positive if they are from the same C / CV category, and negative if the pair from different categories.
A lower EER indicates the system achieves less misclassification in both TD and atypical speech.
On the other hand, 
the AUC measures the overall performance of a binary classifier operating at varying decision thresholds. It is computed from the curve of FPR vs. true positive rate (TPR). 
A higher AUC suggests the detection system is capable to achieve higher TPR with across various operating thresholds. In the assessment of SSD, low FPR is important in the first place, since missing any atypical speech is undesirable. 
If the user tunes the threshold to suppress the FPR, a detection system of high AUC can still reliably classify the TD speech.

\section{Results and Discussion}

\begin{table}[t!]
\centering
\caption{Detection performance using different embeddings.}
\label{hoslistic}
\begin{tabular}{|c|c|c|c|c|}
\hline
\textbf{\begin{tabular}[c]{@{}c@{}}Training\\ Data\end{tabular}}                  & \textbf{Type} & \textbf{Weight} & \textbf{EER} & \textbf{AUC}   \\ \hline
\multirow{3}{*}{\textbf{Child}}                                                   & C & $\lambda$ = 0.9             & 0.142           & 0.918          \\ \cline{2-5} 
                                                                                  & CV   & $\lambda$ = 0.2             & 0.153           & 0.909          \\ \cline{2-5} 
                                                                                  & C+CV & $w$ = 0.4             & 0.120           & 0.932          \\ \hline
\multirow{3}{*}{\textbf{\begin{tabular}[c]{@{}c@{}}Child +\\ Adult\end{tabular}}} & C    & $\lambda$ = 1.0             & 0.126           & 0.935          \\ \cline{2-5} 
                                                                                  & CV   & $\lambda$ = 0.1             & 0.129           & 0.928          \\ \cline{2-5} 
                                                                                  & C+CV & $w$ = 0.5    & \textbf{0.109}  & \textbf{0.945} \\ \hline
\end{tabular}
\end{table}

\begin{figure}[t!]
  \setlength\belowcaptionskip{-0.8\baselineskip}
  \centering
  \includegraphics[width=\linewidth]{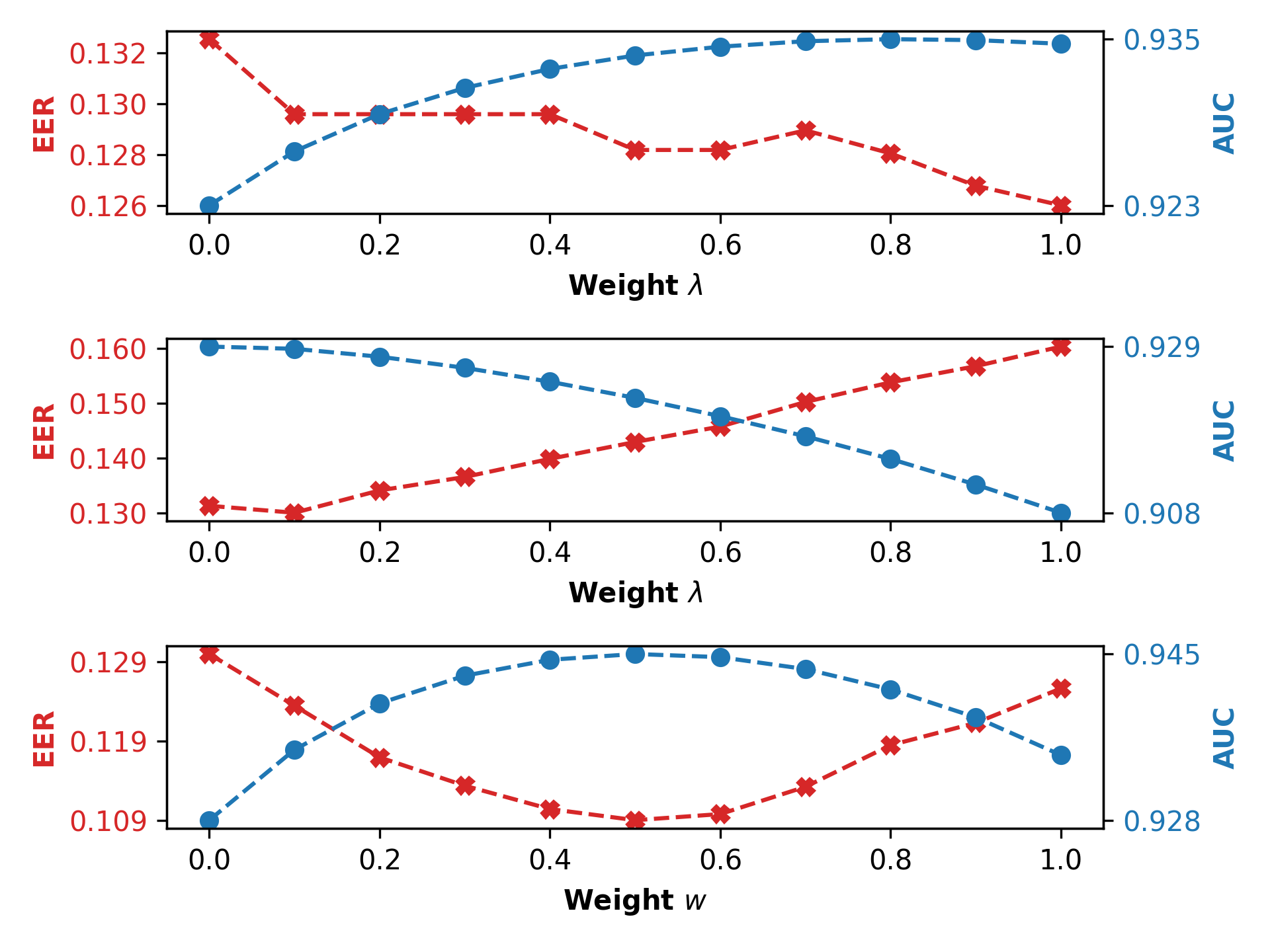}
  \caption{Score fusion vs. detection performance. Top, middle and bottoms figures denote the performance using the embeddings of C, CV, both C and CV, respectively.}
  \label{fig:score fusion}
\end{figure}

\begin{table*}[h!]
\centering
\caption{Detection performance on individual target consonants. \textbf{Bold} indicates best result.}
\label{patterns}
\begin{tabular}{|c|c|c|c|l|c|c|l|c|}
\hline
\multirow{3}{*}{\textbf{Consonant}}      & \multicolumn{2}{c|}{\textbf{Number}}                              & \multicolumn{3}{c|}{\multirow{2}{*}{\textbf{EER}}}                        & \multicolumn{3}{c|}{\multirow{2}{*}{\textbf{AUC}}}                        \\ \cline{2-3}
                                         & \multirow{2}{*}{\textbf{TD}} & \multirow{2}{*}{\textbf{Atypical}} & \multicolumn{3}{c|}{} & \multicolumn{3}{c|}{} \\ \cline{4-9} 
                                         &                              &                                    & C            & \multicolumn{1}{c|}{CV} & C+CV  & C            & \multicolumn{1}{c|}{CV} & C+CV  \\ \hline
/f/                                      & 205                          & 75                                 & 0.079                 & 0.105                            & \textbf{0.077} & \textbf{0.967}        & 0.946                            & 0.965          \\ \hline
/k/                                      & 261                          & 65                                 & 0.196                 & 0.176                            & \textbf{0.154} & 0.896                 & 0.915                            & \textbf{0.925} \\ \hline
/k\textsuperscript{h}/  & 204                          & 160                                & 0.168                 & 0.160                            & \textbf{0.137} & 0.904                 & 0.907                            & \textbf{0.925} \\ \hline
/k\textsuperscript{wh}/ & 66                           & 41                                 & 0.161                 & \textbf{0.086}                   & 0.094          & 0.916                 & \textbf{0.972}                   & 0.956          \\ \hline
/k\textsuperscript{w}/  & 133                          & 23                                 & 0.110                 & 0.129                            & \textbf{0.085} & 0.954                 & 0.918                            & \textbf{0.958} \\ \hline
/l/                                      & 204                          & 36                                 & 0.183                 & \textbf{0.138}                   & \textbf{0.138} & 0.872                 & 0.876                            & \textbf{0.911} \\ \hline
/p\textsuperscript{h}/  & 145                          & 93                                 & 0.118                 & 0.118                            & \textbf{0.097} & \textbf{0.950}        & 0.929                            & 0.946          \\ \hline
/s/                                      & 317                          & 222                                & \textbf{0.084}        & 0.130                            & 0.093          & \textbf{0.956}        & 0.925                            & 0.951          \\ \hline
/t/                                      & 211                          & 63                                 & 0.220                 & \textbf{0.158}                   & 0.173          & 0.844                 & \textbf{0.887}                   & 0.871          \\ \hline
/t\textsuperscript{h}/  & 267                          & 240                                & \textbf{0.116}        & 0.170                            & 0.120          & \textbf{0.943}        & 0.896                            & 0.942          \\ \hline
/ts/                                     & 200                          & 101                                & 0.107                 & \textbf{0.097}                   & \textbf{0.097} & 0.949                 & 0.951                            & \textbf{0.961} \\ \hline
/ts\textsuperscript{h}/ & 243                          & 235                                & 0.067                 & 0.075                            & \textbf{0.050} & 0.974                 & 0.967                            & \textbf{0.975} \\ \hline
\end{tabular}
\end{table*}




In the consonant error detection, each segment from TD and atypical speech is paired up with other TD test segments of the same category. The efficacy is compared between C and CV embeddings. 
For each type of embedding, we find the best combination of the two similarity scores of $score_{cos}$ and $score_{binary}$. In addition, we evaluate the detection performance of combining $score_{C}$ and $score_{CV}$ computed from each test word. 
%
The joint use of adult and child speech in the model training is also validated. The performance of binary detection on all test data is reported in Table \ref{hoslistic}, and Figure \ref{fig:score fusion} illustrates how performance of the extractors change as the weights $\lambda$ and $w$ shift from 0 to 1 in the step-wise manner.

Despite the acoustic mismatches between adult and child speech due to the physiological differences,
the experimental results suggest the direct mixing of adult and child speech is able to boost the detection performance without the use of transfer learning techniques \cite{shivakumar2020transfer}\cite{Qian+2016}. 
It is noted that C embeddings outperforms  CV embeddings as they are used independently in the detection. 

By varying $\lambda$, we observe the detection using CV embeddings relies more on $score_{binary}$. 
C and CV embedding extractors are trained using different numbers of classification targets (19 consonants vs. 173 CV units).
Given the same amount of C and CV segments, each output class of the CV embedding extractor is trained with less speech segments.
On the contrary, the binary relation classifier with single output node is trained by all training segments.
We reckon the insufficient data in the training of multi-class CV classification is likely to make the embedding focus on more the binary relation instead of the global relationship between different CV units.
As the two similarity scores of $score_{C}$ and $score_{CV}$ are combined, best performance is delivered with $0.109$ EER and $0.945$ AUC, where the C and CV extractors are equally contributing to the final score with $w$ equal to $0.5$. 




It remains questionable whether CV embeddings are consistently surpassed by C embeddings across different consonants, and how score fusion improves the detection performance.
The test speech segments are thus divided into $12$ subsets by consonants, and the detection performance is evaluated on individual subset.
The available TD and atypical test segments in each subset and the detection performance are reported in Table \ref{patterns}. 
Using CV embeddings achieves lower EER and higher AUC in the detection of consonant errors in /k/, /k\textsuperscript{h}/,  /k\textsuperscript{wh}/, /l/, /t/ and /ts/. 
The results suggest that CV embeddings deliver a more satisfactory performance in detecting the errors in unaspirated stops of /k/ and /t/. These are the consonants not detected reliably in the previous studies \cite{Ng2020,Wang2019}.
As the C and CV similarity scores of are combined, the system yields the best EER and AUC in most target consonants.
The combined use of C and CV embeddings is able to amend the misclassification caused by either one type of embedding.
In addition, we manually examine the misclassified test segments of which the score fusion does not help.
It is found that some misclassifications is caused by the misalignment of segments, i.e. the consonant is aligned to the vowel segment or background noise. This gives unreliable measurement of similarity scores and harms the detection outcomes.



\section{Conclusion}
We propose and demonstrate the use of consonant-vowel segments in automatically detecting consonant errors in disordered speech at the embedding level.
It has been shown that using consonant-vowel segments improves the detection performance on the challenging unaspirated stop consonants. The approach also achieves comparable performance on detecting other consonants compared to the conventional approach relying on consonant segments.
Calibration of similarity scores computed from both consonant and consonant-vowel segments are investigated. The score fusion is shown to improve the performance across most of the target consonants. Besides, direct mixing of adult speech in training of embedding extractor is able to boost the detection performance. 
Future works include the in-depth acoustical analysis of consonant-vowel interaction in child speech and the subject-level detection of disordered speech.

\section{Acknowledgements}\label{ack}
\vspace{0.5mm}
This research was partially supported by a direct grant and a Research Sustainability Fund from the Research Committee of the Chinese University of Hong Kong, as well as the financial support by the Hear Talk Foundation under the project titled "Speech Analysis for Cantonese Speaking Children".  

\bibliographystyle{IEEEtran}

\bibliography{mybib}


\end{document}